\documentstyle[prl,floats,aps]{revtex}

\begin{document}
\draft

\twocolumn
\narrowtext
\noindent {\bf Comment on "High Field Quasiparticle Tunneling in 
Bi$_2$Sr$_2$CaCu$_2$O$_{8+\delta}$ : Negative Magnetoresistance in the 
Superconducting State"}
 
High magnetic field studies of the cuprate superconductors have 
revealed a non-Fermi liquid origin of the normal state 
resistivity \cite{and}, a negative normal state c-axis magnetoresistance 
\cite{and,zav,zav2}, and  a non-BCS  shape of the upper 
critical field $H_{c2}(T)$ \cite{zav2,alezav,zha}.

However, recently Morozov $et$ $al$ claimed \cite{mor} that Ref. 
\cite{and,zav,alezav} 'fell into the trap' of  attributing the negative c-axis  
magnetoresistance above the maximum of $\rho_c(B)$ to the normal state. 
These authors argued  that magnetoresistance in 
Bi$_{2}$Sr$_{2}$CaCu$_2$O$_{8+\delta}$ is 'controlled by the quasiparticle 
tunnelling in the {\it superconducting} state'.  In this Comment we 
  challenge their arguments.

The authors of Ref.\cite{mor} claim that while $\rho_c$ is a measure
of the 
interplane tunnelling, only the in-plane data could represent a true normal state and 
 be used in the  determination of $H_{c2}$. Their main argument comes 
from the apparently different field dependences of $\rho_c$ and
$\rho_{ab}$ as shown in their Fig.2. 
However, $'\rho_{ab} (B)'$ was measured by means of  
contacts situated on the same face of the crystal \cite{morpr}. Therefore 
this curve does {\it not} represent the in-plane resistivity  because 
neither current redistribution (as considered in \cite{bush}) nor crystal 
imperfections were accounted for.  We have applied the routine 
procedure for resistive $H_{c2}$ evaluation \cite{and,alezav} from the in- 
and out-of-plane data obtained on the {\it same} samples in \cite{and,zha}. 
The very similar values of $H_{c2}(T)$ obtained from $\rho_{ab}$ $and$ $\rho_c$ 
(solid and dashed  lines in the Figure) are evident. This result contradicts the
argument by Morozov $et$ $al$\cite{mor}.  As to  Ref.[11] of \cite{mor}, 
different properties of the crystals used for $\rho_c$ and $\rho_{ab}$ 
measurements (i.e. 10-15K difference in $T_{c0}$ as seen from Fig.3,4 
of that reference), makes corresponding comparison meaningless. 

Another  finding of Ref. \cite{mor} is  a  non-linear c-axis I(V),
attributed by the authors to a suppression of the Josephson coupling.   However, 
a comparative analysis of the data from their Fig.2 and the inset to
Fig.1  favours the Joule heating origin of the effect giving a 
selfconsistent estimate of an excessive heating of about  2K near the
maximum  of the 35K-curve in Fig.3. Moreover, 
although the pulse system used by them delivers 10-20 times smaller rates, 
$dB/dt$, than   conventional systems, the  huge 
size, $S$, of the crystal \cite{mor} essentially overwhelmes this advantage 
since the eddy current heating  is proportional to $(dB/dt)^2S^2$.  
Noticeable heating with eddy currents in the crystal with a 60 times 
smaller $S$ than in Ref. \cite{mor} was observed \cite{zav} below 
30-40\,K.  That suggests significant heating effects in the low-temperature 
data of Ref. \cite{mor}. Indeed, clear signatures of this heating are 
evident from the comparison of their data for 50K and 55K in Fig.1 and 2
 \cite{mor}. An {\it independent} test of the field dependence of the 
quasiparticle conductivity, shown in Fig.4 \cite{mor}, suffers from the 
same problem.  The scale of I(V) curves in the inset to Fig.4, allows us
to estimate  the Joule heating as $\sim30\mu W$. Hence,
heat in excess of 1000\,W/cm$^2$ was 
conducted through the active $2\mu m^2$ area of the mesa depicted in Fig.4.  
Accounting for the poor out-of-plane thermal conductivity of BSCCO-2212, 
this may entail significant heating of the active area of the structure 
thus rising the question of the physical meaning of the measured 
quantities.  

Finally, a linear negative longitudinal magnetoresistance has been 
measured by us \cite{zav,zav2} both below and $above$ the $zero$ field 
critical temperature, $T_{c0}$.  Therefore, in contrast to \cite{mor},
we consider the 
effect to be an intrinsic $normal$ state property. Our view is also supported by a noticeable difference in the absolute value of the normalised 
magnetoresistance slope obtained on  crystals with 
similar $T_{c0}$.

\noindent {\bf V.N. Zavaritsky}, Kapitza Institute for Physical
problems, Moscow, Russia.

\noindent {\bf M. Springford}, H.H. Wills Physics Laboratory,
  University of Bristol, Bristol, United Kingdom.

\noindent {\bf A.S. Alexandrov}, Department of Physics, Loughborough
University, Loughborough, United Kingdom.

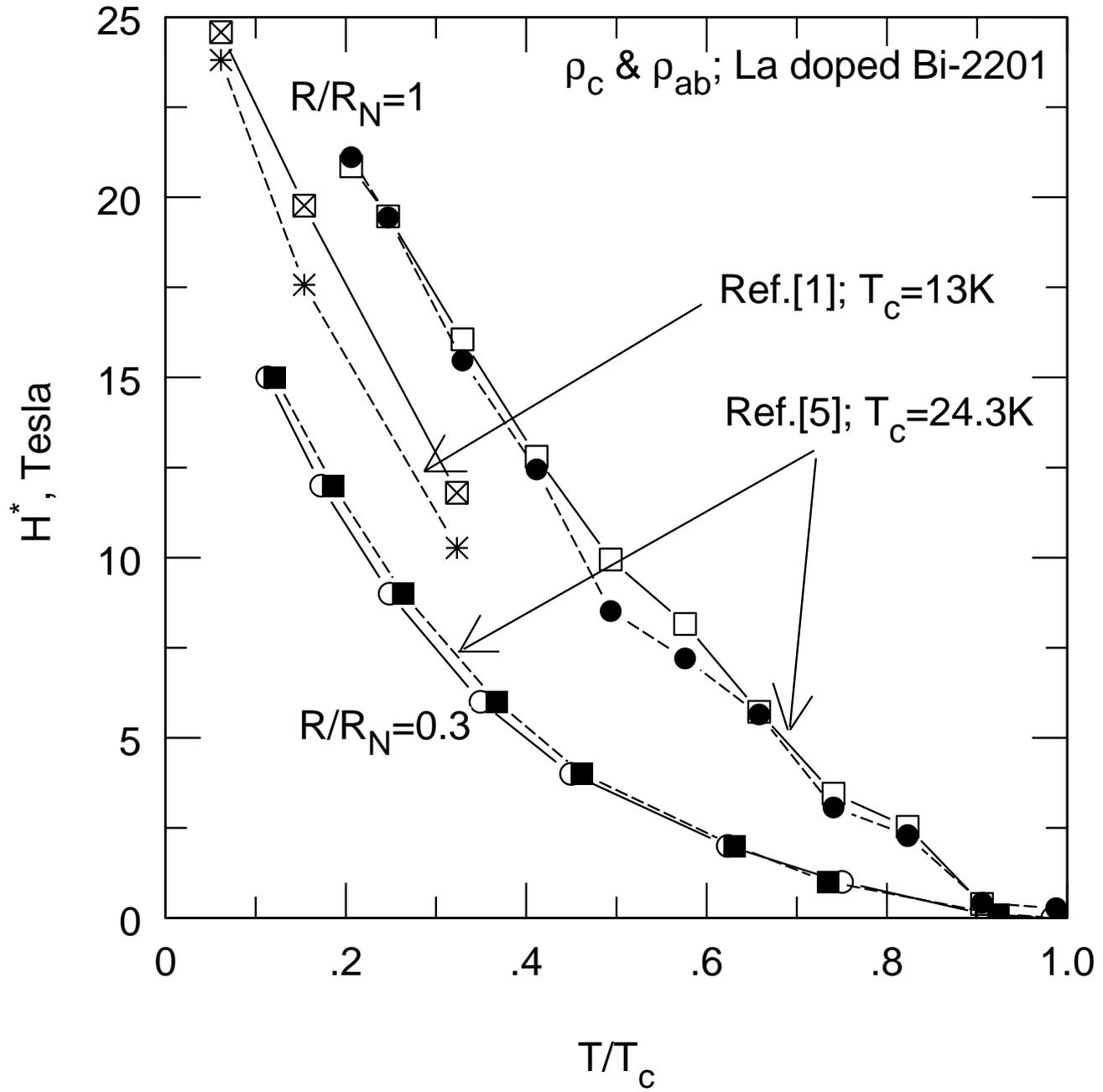
\end{document}